\documentclass[twocolumn,showpacs,preprintnumbers,amsmath,amssymb]{revtex4}

\usepackage{graphicx}
\usepackage{dcolumn}
\usepackage{bm}
\usepackage{nicefrac}

\begin{document}

\preprint{APS/123-QED}

\title{Atomic and molecular phases through attosecond streaking}

\author{Jan Conrad Baggesen and Lars Bojer Madsen}
 \affiliation{Lundbeck Foundation Theoretical Center for Quantum System Research \\
Department of Physics and Astronomy, Aarhus University, DK-8000 Aarhus C, Denmark.}

\date{\today}

\begin{abstract}
In attosecond streaking, an electron is released by a short xuv pulse into a strong near infrared laser field. 
When the laser coupling between two states in the target is weak relative to the detuning, the streaking technique, which allows for a complete determination of the
driving field, also gives an accurate measurement of the relative phase of the atomic or molecular ionization matrix elements from the two states through the interference from the two channels. The interference may change the phase of the photoelectron streaking signal within the envelope of the ir field, an effect to be accounted for when reconstructing short pulses from the photoelectron signal and in attosecond time-resolved measurements.
\end{abstract}

\maketitle

With the ability to control and shape infrared (ir) laser pulses, a number of new posibilities are emerging. The ability to create and characterize intense few-cycle phase stabilized pulses\cite{goulielmakis2004} has allowed the formation of isolated attosecond pulses\cite{kienberger2004,goulielmakis2004,cavalieri2007}, that in combination with the original ir pulse has opened the research field of attoscience\cite{krausz2009}. In attoscience the quest is for measurements with attosecond temporal resolution\cite{cavalieri2007,schultze2010} and simultaneous angstrom spatial resolution, ideally retrieving both amplitude and phase of the atomic and molecular wavefunctions in real time as an electron moves\cite{itatani2004,haessler2010}. 

The retrieval of atomic and molecular phases requires interference between different channels and has recently been pursued in the context of two-photon ionization with a train of attosecond pulses combined with an ir laser pulse\cite{haessler2009,swoboda2010}. In these experiments, the measured atomic phase is the phase of the two-photon matrix elements, which involves a weighted  sum over the complete spectrum of field-free intermediate states. Other works measure multiphoton strong-field ionization phases using the process of high-order harmonic generation~\cite{mairesse2010}. In this work, we show how to measure the phase of the fundamental one-photon ionization matrix element from an initial state to a continuum final state. These atomic or molecular phases may serve as valuable tests for detailed calculations of the ionization process, and
may give direct experimental access to the variation of scattering phase shifts with energy - a quantity determining the Wigner time delay~\cite{schultze2010,Wigner1955}.

We consider a situation,  where the initial state $\vert 1 \rangle$ is coupled weakly by the ir field to 
a detuned excited state $\vert 2 \rangle$ such that only part of the initial population is transferred to $\vert 2 \rangle$ and only while the driving ir pulse is on, adiabatic following the ir pulse envelope. The relative phase between the amplitudes of $\vert 1 \rangle$ and $\vert 2 \rangle$  will be phase-locked to the ir field. We then use an attosecond pulse with a duration much shorter than the optical period of the ir field to ionize from both the levels, and the two contribution will overlap in energy and add coherently in the overlapping region (Fig.~1). We will show that measuring the relative phase of the two contributions and at the same time the ir field through the attosecond streaking process\cite{goulielmakis2004} allows a reconstruction of the relative atomic or molecular phases from the two ionization channels.

\begin{figure}[ht]
	\includegraphics[width=0.30\textwidth]{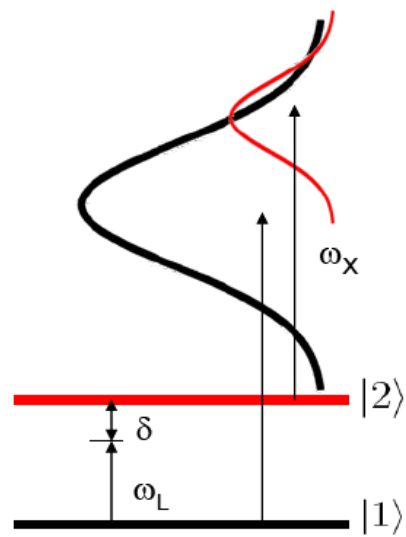}
	
	\caption{(Color online) Schematic view of the ionization process. The two levels $\vert 1 \rangle$ and $\vert 2 \rangle$ are coupled by the ir laser field, and the spectral width of the attosecond pulse is sufficiently large that photoelectrons  from both levels overlap energetically. The interference between the two ionization channels in the overlapping region carries a signature of the phase between the ionization matrix elements.}
	\label{introduction}
\end{figure}

The two-level system is described by the wavefunction [atomic units (a.u.) with $\hbar = e = a_0 = m_e = 1$ are used throughout unless indicated otherwise]
\begin{equation}
	\Psi(t) = c_1(t)e^{-iE_1t}|1\rangle + c_2(t)e^{-iE_2t}|2\rangle
	\label{eq:psit}
\end{equation}
with the initial conditions $c_1(-\infty)=1$, $c_2(-\infty) = 0$. The total Hamiltonian of the system is $H = H_0 + V_L(t)$, where the two levels are Eigenstates of the field-free Hamiltonian $H_0$ and $V_L$ is the interaction of the system with the ir laser field. The coefficients $c_i(t)$ may be found by solving the set of coupled differential equations
\begin{eqnarray}\label{eq:diffequ}
	\dot c_1(t) &=& -i\langle 1|V_L(t)|2\rangle e^{i(E_1-E_2)t}c_2(t) \\
	\dot c_2(t) &=& -i\langle 2|V_L(t)|1\rangle e^{i(E_2-E_1)t}c_1(t)\nonumber.
	\end{eqnarray}
Analytical approximations may be applied to these equations, giving interesting insights that will be detailed elsewhere. In this work, we parametrize the interaction as $\langle 1|V_L(t)|2\rangle = \rho F_L(t)$, where $\rho$ is the coupling strength and $F_L(t)$ is the electric field strength of the ir field. We then solve the equations numerically.

We are interested in a regime with  large detuning, $\delta = E_2-E_1-\omega_L > \rho F_L$, where $\omega_L$ is the central laser frequency. In this regime, there are no Rabi oscillations and the population only follows the field adiabatically. After the pulse, virtually all of the population is left in the ground state again. When the coupling is much smaller than the detuning, the solution to \eqref{eq:diffequ} may be approximated with $c_1(t)\sim 1$ and $c_2(t) \sim -\nicefrac{\rho F_L^0(t)e^{i\delta t}}{2\delta}$, where $F_L^0(t)$ is the envelope of the electric field, such that $F_L(t) = F_L^0(t)\cos(\omega_L t + \phi)$. In this regime, the phase-evolution of the two terms in \eqref{eq:psit} is determined by the laser frequency and the amplitude in the upper level follows the envelope $F_L^0(t)$.

We now add a second, weaker xuv attosecond pulse which ionizes from the state \eqref{eq:psit}. The duration of the xuv pulse $\tau_x \ll \omega_L^{-1}$ is sufficiently short that all of the electrons will be released at a well-defined phase of the ir field. We vary the time, $t_0$, of the xuv pulse relative to the ir pulse and calculate the electron distribution in first order perturbation theory through $\nicefrac{dP}{d\vec k_f} = |T_{fi}|^2$, with
\begin{equation}
	T_{fi}(t_0) = -i \int dt \langle k_f(t)|V_x(t-t_0)|\Psi(t)\rangle.
	\label{eq:Tmat}
\end{equation} 
Here, we approximate the time-dependence of the final state with the Volkov-phase, such that $|\vec k(t)\rangle = |\vec k\rangle e^{-\frac{i}{2}\int^t dt'(\vec k + \vec A(t'))^2}$, where the vector potential is given by $\vec F_L(t) = -\nicefrac{d\vec A}{dt}$ and $\vec k$ is the asymptotic momentum of the scattering state. 
In the strong-field approximation $\vert \vec k\rangle$ is approximated by a plane wave, neglecting the Coulomb potential in the final state, but this is 
not necessary for the present theory.
 We factorize the matrix elements as $\langle k|V_x(t-t_0)|n\rangle = M_n(\vec k) F_x(t-t_0)$, where $n = 1,2$ from \eqref{eq:psit} and $M_n(\vec k)$ is the ionization matrix element for going from the state $n$ to the final state $|\vec k\rangle$, which we wish to determine. With this factorization and using \eqref{eq:psit} we have $\langle k)|V_x(t-t_0)|\Psi(t)\rangle = \sum_{n=1,2}c_n(t)M_n(\vec k)F_x(t-t_0)$.

The attosecond xuv pulse has a Gaussian envelope and is given by $F_x(t) = F_{x,0}e^{-\frac{t^2(1+i\xi)}{\tau_x^2(1+\xi^2)}}e^{-i\omega_xt}$, where $\omega_x$ the central frequency and $\xi$ is the dimensionless chirp of the pulse, which can be measured with the streaking technique\cite{itatani2002}. The ir pulse has a sine-squared envelope, $F_L(t) = \sin^2\left( \frac{\pi t}{\tau_L}\right)\cos(\omega_Lt+\phi)$, which allows the integrals in the Volkov phase to be calculated analytically. 

\begin{figure}[ht]
	\includegraphics[width =0.45\textwidth]{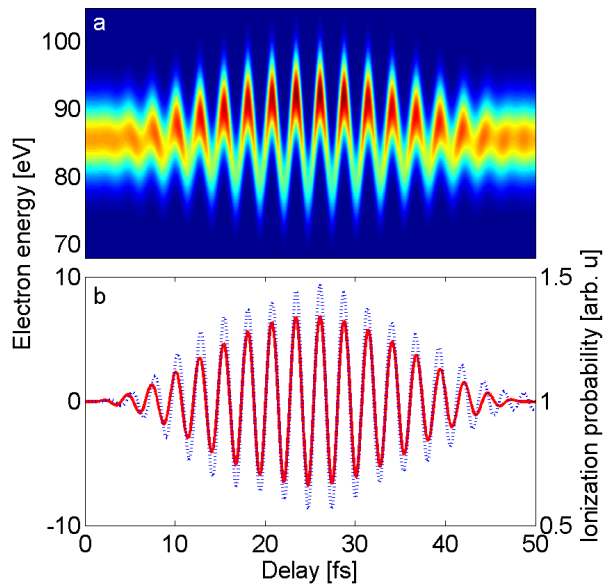}
	
	\caption{(Color online) (a) Photoelectron spectrum as a function of time delay between the ir and the attosecond pulse, from the laser-driven two-level system discussed in the text. (b) The center of energy (full, red) and the energy integrated ionization yield (dashed, blue). The two curves are oscillating in phase as a signature of the $\nicefrac{\pi}{2}$ phase difference between the two matrix elements $M_1$ and $M_2$.}
	\label{fig1}
\end{figure}

As an example, we calculate the ionization spectrum in an atom with $\rho = a_0$ and energy difference of 1.85 eV corresponding to the energy of the first excited state in lithium. The ionization potential of the ground state is 5.39 eV. For the (near) ir field we choose a 800 nm,  50 fs pulse with  peak intensity $1\times 10^{12}$ W/cm$^2$. The attosecond pulse has a full width at half maximum duration of 290 as and a central frequency corresponding to  91 eV. We assume that the electrons are detected parallel to the polarization of the ir field. In the first example, the two atomic ionization matrix elements are chosen to be  ${\pi}/{2}$ out of phase, such that $M_2(\vec k) = iM_1(\vec k)$. If the final state $\vert \vec k \rangle$  is assumed to be a plane wave, this would be the relation between two states of opposite inversion symmetry along the line of detection. Since the two contributions are from different initial states, they do not have the same symmetry in the ionization continuum. Any phase difference would manifest itself as a deviation from the $\nicefrac{\pi}{2}$ phase difference between the matrix elements. The results are presented in Fig.~\ref{fig1}.

\begin{figure}[ht]
	\includegraphics[width =0.30\textwidth]{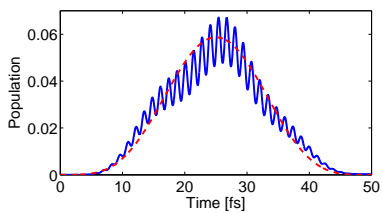}
	
	\caption{(Color online) The numerically calculated population in the excited state, $|c_2(t)|^2$ (full, blue) and the approximate result from $c_2(t) \sim -\nicefrac{\rho F_L^0(t)e^{i\delta t}}{2\delta}$ (dashed, red).}
	\label{population}
\end{figure}

The population transfer with the given set of parameters is always less than 7\%, but still the ionization yield oscillates by around 40\%. In the limit of weak coupling the amplitude of the oscillation is given by $\nicefrac{\rho F_L^0(t)}{\delta}$. In Fig.~\ref{population}, the population in the excited state is shown along with the weak limit estimate. The population left in the excited state after the pulse  is of the order of $10^{-4}$, such that the target is left in the ground state. When the laser coupling between the two states is smaller than the detuning, such that there are no Rabi oscillations, the phase evolution of the excited level will be locked to the laser pulse. This is seen in Fig.~\ref{fig1} (a) by the fact that the maxima in the emission probability coincide with the maxima in the electron energy, which follows the vector potential of the ir field through the classical electron acceleration $\Delta \vec k = -\int_{t_0}^\infty \vec F_L(t)dt = -\vec A(t_0)$ and $\Delta E(t_0) \sim -\vec k\cdot \vec A(t_0)$. Since the energy-integrated ionization probability is determined by the relative phase of the two contributions originating from the ground state and the excited state, the ionization probability will be locked to the laser field. The advantage of the attosecond streaking technique is, that it allows the precise determination of the phase of the ir field\cite{goulielmakis2004,mairesse2005}. Hence, as we can simultanously measure the phase of the ir field and the relative phase of the electrons ionized from the two levels, we can determine the relative phase of the ionization matrix elements $M_1$ and $M_2$. 

\begin{figure}[ht]
	\includegraphics[width =0.45\textwidth]{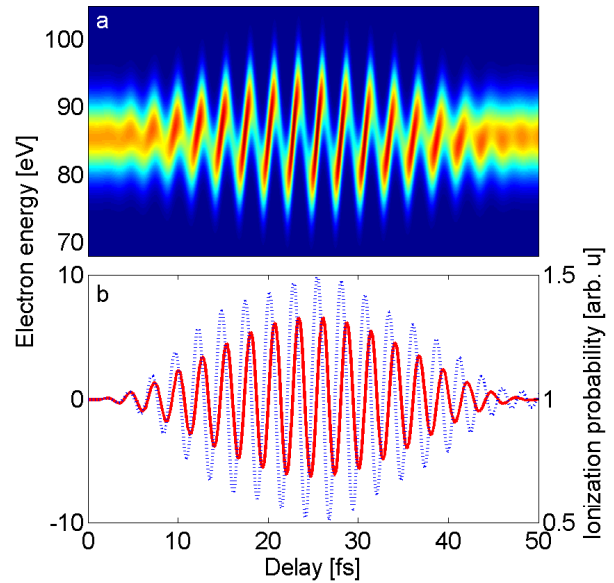}
		
	\caption{(Color online) (a) Photoelectron spectrum from the laser driven two-level system discussed in the text, similar to Fig~\ref{fig1}, but with the two matrix elements in phase. (b) The center of energy (full, red) and the energy integrated ionization yield (dashed, blue). The two curves are oscillating $\frac{\pi}{2}$ out of phase as a signature of the two matrix elements $M_1$ and $M_2$ being in phase.}
	\label{fig2}
\end{figure}

For the same range of parameters as above, but with the two ionization matrix elements in phase, $M_1=M_2$, the photoelectron spectrum is plotted in Fig.~\ref{fig2}. As can immediately be seen from the photoelectron spectrum, the maxima of the ionization probability now coincide with half the zero-crossings of the vector potential, showing that the relative phase between the vector potential and the ionization yield is a strong measure of the relative phase between the ionization matrix elements. This is more clear when looking at the center of energy along with the energy integrated ionization probability in Figs.~\ref{fig1} and \ref{fig2} (b). In Fig.~\ref{fig1}, the two curves oscillate in phase, while in Fig.~\ref{fig2} the two curves oscillate $\nicefrac{\pi}{2}$ out of phase; the difference in the phase of the oscillations maps directly to the difference in the relative phase of the ionization matrix elements.

All of this applies equally well to atoms and molecules, as long as there is a coupling between the ground state and an excited state. To accurately retrieve the atomic or molecular phases, it is necessary to determine the precise phase of the vector potential. For polar molecules, the center of energy of the photoelectrons does not exactly follow the vector potential, but may be shifted due to the linear Stark shift of the initial state\cite{baggesen2010}. In this case, in order to accurately retrieve the phases of the ionization matrix elements, either a nonpolar reference state or knowledge of the dipole moments is needed in order to correct for these effects. 

\begin{figure}[ht]
\includegraphics[width =0.45\textwidth]{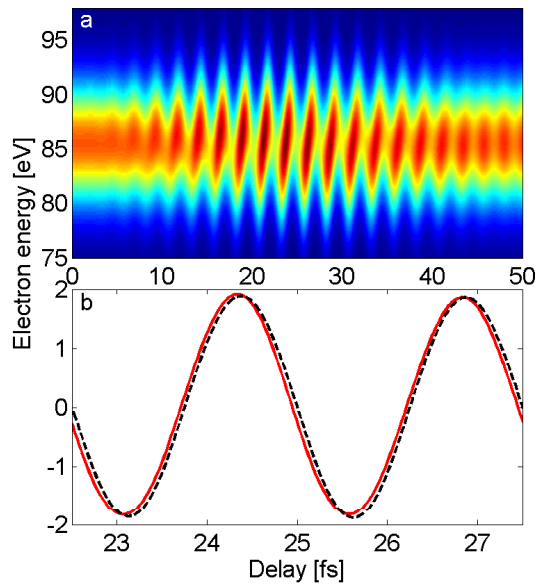}
	
	\caption{(Color online) (a) Photoelectron spectrum from the laser driven two level system, parameters are presented in the text. (b) Part of the center of energy (full, red) is plotted along with the classical estimate of the energy variation (dashed, black). The two curves are displaced by 50 as or 0.13 rad}
	\label{fig3}
\end{figure}

Effects similar to the polarization effects\cite{baggesen2010} may arise from the initial state \eqref{eq:psit}. If the ionization probability and the vector potential are not exactly in phase, that is, when the ionization matrix elements are not $\nicefrac{\pi}{2}$ out of phase, then the difference in energy of the two contributions leads to a shift in the center of energy of the electrons. Since the dominant contribution is from the ground state, a constructive interference will lead to an increase of the average electron energy while a destructive interference will lead to a decrease of the electron energy. Now, as the interference is phase-locked to the driving laser field, this means that the average electron energy will be changed from the classical estimate $\Delta E(t_0) = -\vec k\cdot \vec A(t_0)$ and the electron energy is not a definate measure of the phase of the vector potential. This may be seen in Fig.~\ref{fig3}, where the center of energy is plotted along with the classical estimate for the same system as in Fig.~\ref{fig2}, but for a laser intensity of $1\times 10^{11}$W/cm$^2$ and with a laser wavelength of 750 nm, closer to the resonance.

In this example a shift of 0.13 rad or 50 as is found. We find that in the limit of large detunings, the phase shift of the electrons is not sensitive to the intensity of the laser. In a simple model for weak couplings to be detailed elsewhere the phase shift in radians may be estimated as 
\begin{equation}
\Delta=\frac{\rho\omega_L^2}{2\delta \vec k\cdot \vec \epsilon_L}, 
\end{equation}
where $\epsilon_L$ is the laser polarization vector. For the range of parameters used above, this estimates the phase shift to be $\simeq 0.11$ rad corresponding to a time shift of $\Delta/\omega_L =42$ as.

That fact that the center of energy of the photoelectrons is not exactly in phase with the vector potential may be important when the attosecond streaking technique is used both to retrieve atomic or molecular phases and when it is used for time-resolved measurements. It may give rise to an apparent shift between valence electrons where near resonances are present and core electrons with no resonances. The time delay of 50 as in the example above is comparable to the temporal resolution one can achieve\cite{cavalieri2007,schultze2010} and must therefore be taken into consideration.  These shifts may be circumvented either by calculating the phase shift in the photoelectrons and correcting for it or by using photoelectrons from a different level to determine the phase of the vector potential. This level should have no near resonances and be nonpolar to most easily extract the vector potential.

In scattering theory, a $k$ gradient of the scattering phase is known to give rise to a Wigner time-delay\cite{Wigner1955}, which may manifest itself as a time delay in attosecond streaking experiments~\cite{schultze2010}. The Wigner time delay is a purely phase-dependent effect and detailed knowledge of the phase of the ionization matrix element may lead to a better understanding of the importance of this effect.

In conclusion, we have shown that when an ir pulse with a large detuning excites a two-level system, the phase evolution of the system will be locked to the pulse and the ionization probability will oscillate with the driving frequency. The phase of this oscillation is determined by the phase of the driving laser field and the phase of the atomic or molecular ionization matrix elements. With the attosecond streaking technique, it is posible to extract the phase of the driving field and hence use these oscillations to measure the atomic or molecular phases which have so far resisted experimental investigation.



\end{document}